# Detecting the historical roots of research fields

# by reference publication year spectroscopy (RPYS)

Werner Marx[1], Lutz Bornmann[2], Andreas Barth[3], Loet Leydesdorff[4]

[1] w.marx@fkf.mpg.de

Max Planck Institute for Solid State Research, Heisenbergstraße 1, D-70569 Stuttgart, Germany.

Corresponding author

[2] bornmann@gv.mpg.de

Division for Science and Innovation Studies, Administrative Headquarters of the Max Planck Society,

Hofgartenstraße 8, 80539 Munich, Germany.

[3] andreas.barth@fiz-karlsruhe.de

FIZ Karlsruhe, Hermann-von-Helmholtz-Platz 1, D-76344 Eggenstein-Leopoldshafen, Germany.

[4] loet@leydesdorff.net

Amsterdam School of Communication Research (ASCoR), University of Amsterdam, Kloveniersburgwal 48, 1012

CX Amsterdam, The Netherlands.

**Abstract**

We introduce the quantitative method named "reference publication year spectroscopy" (RPYS). With this method one can determine the historical roots of research fields and quantify their impact on current research. RPYS is based on the analysis of the frequency with which references are cited in the publications of a specific research field in terms of the publication years of these cited references. The origins show up in the form of more or less pronounced peaks mostly caused by individual publications which are cited particularly frequently. In this study, we use research on graphene and on solar cells to illustrate how RPYS functions, and what results it can deliver.

**Introduction**

Research activity usually evolves on the basis of previous investigations and discussions among the experts in a scientific community: "Original ideas seldom come entirely 'out of the blue'. They are typically novel combinations of existing ideas" (Ziman, 2000, p. 212). Earlier findings are re-combined and developed further on, resulting in the accumulation of knowledge and thus in scientific progress. According to Popper (1961) scientists formulate empirically falsifiable hypotheses, develop empirical tests for these hypotheses, and apply them. Some hypotheses are corroborated as this process is repeated or applied in different contexts and others are rejected. Thus, knowledge is acquired when hypotheses are formed on the basis of earlier findings and by the empirical testing they undergo. In Kuhn's (1962) alternative view, knowledge is acquired when scientists work on specific problems or puzzles. According to Kuhn (1962) scientists working under normal circumstances are guided by paradigms or exemplars which provide a framework for the work (puzzle-solving). Paradigms are "a set of guiding concepts, theories and methods on which most members of the relevant community agree" (Kaiser, 2012, p. 166). When scientists question what represents good evidence and reason in a research field, and a different framework offers a better alternative, one paradigm replaces the other. Kuhn therefore believes that knowledge is acquired through changes in paradigms in a non-cumulative process. Popper (1961), on the other hand, envisages a cumulative process: "Popper is more concerned with the normative and prescriptive question of how science *should* be carried out, and Kuhn is more concerned with the descriptive question of how science *is* carried out" (Feist, 2006, p. 30).

Although there are many differences between these two theories of scientific development, the relationship of current research to past literature plays a significant role in both: knowledge cannot be acquired without this relationship. The relationship to earlier publications is expressed in the form of references. One can expect that the content of an earlier publication and that of the later publication are related and that the former is of significance to the knowledge claim in the latter. The premise of a normative theory of citations and its application to the evaluation of research is that, in terms of statistics, the more frequently scientific publications are cited, the more important they are for the advancement of knowledge (Merton, 1965; Bornmann et al., 2010). From this perspective, citation data provides interesting insights into the historical science context, in terms of the significance of the previous publications on which the later publications in a field of research are based.

In this study we introduce a quantitative method to reveal the most important historical publications of a specific research field based on the analysis of the publication years of the references cited within the relevant literature. The focus on the important historical publications in a specific research field is a special application of the method known as cited reference analysis (Bornmann & Marx, 2013). In an analogy to the spectra in the natural sciences, which are characterized by pronounced peaks in the quantification of certain properties (such as the absorption or reflection of light as a function of its color), we call this special application "reference publication year spectroscopy" (RPYS) (Marx et al., 2013). To illustrate RPYS we show two examples of how it is possible to determine and further analyze the historical roots of the publications cited within a specific research field: research on graphene and on solar cells and photovoltaics, respectively.

The analysis of the reference publication years is not a new bibliometric approach but has already been discussed by De Solla Price (1974; p. 91). And it was for instance applied by Van Raan (2000) to measure the growth of science and to detect important breakthroughs in science without pre-defining any field. The method presented here is also based on the citation-assisted background (CAB) method proposed by Kostoff and Shlesinger (2005) which is a "systematic approach for identifying seminal references" (p. 199) in a specific field (Kostoff et al., 2006).

**Methods**

Citation analyses are usually based on document sets comprising the publications of a researcher, a research institution, or published in one or more specific journals. The number of times these publications are cited is analyzed to evaluate research performance. As a rule, citations from every research field and not only those of the citing publications within a certain research field are taken into account. In a previous publication (Bornmann & Marx, 2013), rather than starting with citations of the publications in one research field, for certain issues we proposed reversing the perspective and analyzing the publications referenced in publications in the same research field in order to determine the impact of publications, authors, institutions or journals within this specific research field.[1] We have shown that it is possible to limit citation analyses to single research fields by first selecting their publications and then analyzing the references cited in them. A cited reference analysis with specific emphasis on the publication years of the references can be used to quantify the significance of historical publications and to reveal the historical roots of a given research field.

Empirically, it has been shown that most references refer to more recent specialist literature in the discipline in which the citing publication has appeared – only a relatively small proportion of the cited publications is older and derives from different disciplines. The distribution of the cited publications over their publication years (the reference publication years, RPYs; not to confuse with the method RPYS) is typically at the maximum a few years before the publication year of the citing publications and then tails off significantly into the past. The (steep) decline over time is not only associated with the fact that specialist literature as a rule becomes less interesting and important as time passes (ageing). It is also the result of an abrupt increase in specialist literature in every discipline which began around 1960 ("Sputnik shock") and continues to this day. For example, just 2% of the literature on physics in the 20th century was published before 1950 (Marx, 2011).

Quantitative analysis of the publication years of all the publications cited in the publications in one specific research field shows that RPYs lying further back in the past are not represented equally, but that some RPYs appear particularly frequently in the references. These frequently occurring RPYs become more differentiated towards the past and mostly show up as distinct peaks in the RPY distribution curves. If one analyses the publications underlying these peaks, it is possible to see that during the 19th and the first half of the 20th century they are predominantly formed by single relatively highly cited publications. These few, particularly frequently cited publications as a rule contain the historical roots to the research field in question. These publications can be found with cited reference analysis (Bornmann & Marx, 2013) and it is possible to determine how the relationship to earlier publications developed over time; that is, at which stage in the development of the research field these publications were (re-)discovered and then cited more frequently. Towards the present, the peaks of individual publications lie over a broad continuum of newer publications and are less pronounced. Due to the many publications cited in the more recent RPYs, the proportion of individual, much-cited publications in the RPYs falls steadily.

---

[1] In another publication, Costas et al. (2012) proposed analyzing the typical use of bibliographical references by individual scientists.

The results of the RPYS method applied on graphene and solar cells literature presented here are based on the Science Citation Index (SCI) which is accessed via the SCISEARCH database offered by the database provider STN International (http://www.stn-international.com/). This database combined with the STN search system enables sophisticated citation analyses. Among many other options, the SCISEARCH database searched via STN International makes it possible to ask which historical publications in the various fields of the natural sciences have been cited most frequently by the publications since 1974, the period covered by the SCISEARCH database. The Web of Science (WoS) provided by Thomson Reuters, the most common search platform of the Thomson Reuters citation indexes, stretches back to 1900. However, the WoS search functions have not been optimized for the bibliometric analysis presented in this study. The selection of numerous references from large sets of citing publications and their further analysis is not possible under WoS without first downloading the citing documents. In the case of RPYS, one has to download the data before further processing, for example, in Excel.

STN's retrieval system allows the publications from a specific research field to be selected and all the references they cite to be extracted. Instead of the complete references it is also possible to select and analyze only the authors of the publications in the cited references, the journals or the RPYs. In this study, we are concerned mainly with the analysis of the RPYs and especially the early publications cited particularly frequently as the historical roots of a research field. The first step in RPYS is to select the publications for a certain research field and extract all the references from them. The second is to establish the distribution of the frequencies of the cited references over the RPYs and from this determine the early RPYs cited rather frequently (a minimum citation count of 10 has proved to be reasonable). The third is to analyze these RPYs for frequently cited historical publications.

Before we describe the results of the RPYS method in the following two sections, we will present for each example an overlay of journals on the basis of aggregated journal-journal citations in order to position the graphene and solar cell research fields within the whole of science. This routine is based on the cosine-normalized citations among journals contained in the Journal Citation Reports (Thomson Reuters) 2011 of the Science and Social Science Citation Index (SCI and SSCI combined). The data is projected onto a two-dimensional map using the software VOSViewer (Van Eck & Waltman, 2010) for the mapping and Blondel et al.'s (2008) community-finding algorithm for the coloring into 12 disciplines. In this study, we follow Leydesdorff et al. (in press) and use the maps normalized on the citing side (journals, not reference journals) of the matrix for the comparison. Rao-Stirling diversity $\Delta$ is used as a measure of interdisciplinarity of the downloaded sets. This measure is defined as follows:

$$\Delta = \sum_{ij} p_i p_j d_{ij}$$

(1)

where (in this case) $p_i$ is the relative frequency of each journal in the document set and $d_{ij}$ the distance between two journals as a fraction of the maximal possible distance on the two-dimensional map generated by VOSViewer.[2]

The Rao-Stirling diversity measure was introduced by Rao (1982a and b) and has also been named "quadratic entropy" (Izsák & Papp, 1995) because it measures not only diversity in terms of the spread of the publications among the journals, but also takes into account the distances among the journals on the map. Stirling (2007, at p. 712) proposed this measure as a general framework for measuring diversity in science, technology, and innovation. Porter et al. (2007) used this measure in their integration score of interdisciplinarity.

**Results**

**Example 1: Graphene**

---

[2] Leydesdorff, Rafols, and Chen (in press) note that the distance between two journals can be calculated in the multidimensional space using $(1 - cosine)$ as a distance measure, but that for reasons of computational efficiency the distance on the map $\|x_i - x_j\|$ can also be used given the spatial reduction of the complexity used by VOSViewer.

Single planar layers of graphite one atom thick are named graphene, the newest member of the carbon structural family. Graphene has been called a rising star among the new materials (Geim & Novoselov, 2007; Barth & Marx, 2008). Although graphene has been discussed since 1947, it was not believed to exist in a free state. In 2004, however, graphene was found unexpectedly when it was isolated from graphite crystals (Novoselov et al., 2004; Novoselov et al., 2005). This defined a new allotrope of carbon in addition to diamond, graphite, nanotubes, and fullerenes. Graphene exhibits some remarkable properties which feature in particular highly efficient electrical conductivity combined with extremely fast charge transport and extraordinary strength. These properties make the material potentially useful in a wide range of applications such as electronics (high speed transistors and single-electron transistors) and materials science (composite materials) (Geim & Novoselov, 2007; Geim & Kim, 2008). The experimental discovery of free-standing graphene sheets as a new member of the carbon structural family caused a "gold rush" to surround this interesting and promising research field, leading to a substantial rise in the number of publications. Since research on graphene has become a "hot topic" for scientists, it is not surprising that the publication (and citation) pattern of such a new research field is also of great interest for scientometric studies (see e.g. Winnink, 2012).

What does it mean: graphene research? For the subject-specific visualization of research on graphene we downloaded the set of papers with the string "TI=graphene" from WoS on February 7, 2013 (see Table 1). The retrieval result was 16,145 records. This search is based on title word searching only (which is sufficient for this overview). The 16,145 records are used as input to the journal mapping using the routine provided by Leydesdorff, Rafols, & Chen (in press). The maps are overlaid on a base map that uses grey dots to indicate the journals which are not covered by the document set under study. In the other case, the logarithm of the number of documents is used for sizing the node. The nodes are colored in accordance with the community-finding algorithm of Blondel et al. (2008) applied to the entire matrix of 10,675 journals covered by the Journal Citation Reports of the Science Citation Index and Social Science Citation Index combined. The twelve colors can thus be considered as an estimate for a disciplinary attribution (Leydesdorff et al., in press).

Insert Table 1 here

Insert Figure 1 here

Figure 1 shows at a glance that the graphene research field stretches throughout large areas of the natural sciences. Research on graphene concerns basic aspects of chemistry and physics: a new carbon modification, a new material with extraordinary properties and potential applications as well as many theoretical considerations. However, the research field has a strong focus on one specific journal: "Physical Review B".

To apply RPYS to the graphene research field, the publications dealing with graphene were selected by searching for the term "graphene" in the title and abstract search fields of the SCISEARCH database (the abstracts have been considered here to ensure completeness). There is no need for a search in a field-specific database such as the Chemical Abstracts Service (CAS) literature database because the core literature covered by SCISEARCH can be expected to be sufficient to reveal the most frequently cited historical publications (Moed, 2005; p. 126-127). The STN search query for the RPYS of the graphene literature is given in Table 2. The selection of the graphene-relevant papers is easily possible with only one specific search term (graphene). Sometimes, however, the field-specific literature is accessible only by more complex search terms which have to be explored and checked carefully.

Of the complete set of 23,443 publications on graphene research (mentioning "graphene" in the title or abstract text) and published since 1974 (the time period covered by the SCI accessible under STN International) in one of the SCI source journals, all cited RPYs (190) have been extracted (date of the literature search: Feb. 07 2013).

Insert Table 2 here

Notes. L1: Selection of the publications dealing with graphene by searching for the term "graphene" in the title and abstract search fields. L2: Extraction of the RPYs from all of the cited references (both list number entries marked in light grey). The number of references with RPYs from 1850 to 1870 (cut-

outs of the full STN-specific display list including the earliest pronounced peak with n=156/102 cited references in 1859/1860, again marked in light grey) are displayed here for demonstration. Source: SCISEARCH under STN International.

The distribution of the number of references cited in graphene literature across the publication years is presented in Figures 2a-2d. Figure 2a shows the distribution of the number of all the references cited in graphene publications across their publication years. The most frequently cited RPY is 2009, showing the strong contemporary relevance of this newly emerging research field. The RPYs are presented here back to the year 1800 – the pre-1800 references are much less numerous, much more erroneous, and also less important, because the corresponding publications appeared prior to "modern science". Figure 2b shows a cut-out limiting the RPYs to 1800-1990 with the distinct peaks of the most frequently cited historical publications more clearly visible. The citing graphene publications were published between 1974 and the present (mainly since 2004), whereas the time window of the cited publications (the references cited within the citing graphene publications analyzed here) extends from 1800 to 1990 in order to focus on historical publications and to provide suitable scaling to reveal the peaks. Figures 2c and 2d show the deviation of the number of cited references in one year from the median for the number of cited references in the two previous, the current and the two following years. While Figure 2c shows the absolute deviation from the 5-year median, Figure 2d illustrates the deviation from the 5-year median in percentage. It is particularly easy to see the peaks created by the frequently cited historical publications in the deviations normalized as a percentage of the cited references in the corresponding 5 years. As mentioned above, the Figures 2b-2d show that the more frequently occurring RPYs become more differentiated towards the past and mostly show up as distinct peaks in the RPY distribution curves.

Inset Figure 2 here

The search query for the citation analysis of the peak in the RPYs 1859/1860 via the SCISEARCH database under STN International is given in Table 3 to illustrate the RPYS database procedure. The 1859/1860 peak is absolutely remarkable since science was small-sized in the 19th century. As the list of references shows, many references have turned out to be erroneous. Misspelled citations (e.g.

incorrect with regard to the numerical data: volume, starting page, and publication year) are a general problem in citation analysis (Leydesdorff, 2008). The references in earlier publications, however, are particularly susceptible to 'mutations' (Marx, 2011).

Insert Table 3 here

Notes. L3-L10: List numbers comprising the search steps of the analysis of the RPYs 1859/1860 with peaks and demonstrating the analysis method by displaying the reference variants of the publications by Brodie (1859 and 1860) as an example (with the relevant search steps marked in light grey). The reference variants mainly result from the missing of full reference journal name standardization and from misspellings. Source: SCISEARCH under STN International.

The RPY peaks arising in the first half of the 20th century are predominantly formed by single relatively highly cited publications. The four most clearly pronounced peaks in Figure 2d can be attributed to early publications on graphite oxide which are most important for graphene research. Table 4 specifies the four most frequently cited historical publications, including their bibliographic data and comments on the publications taken from a review on graphene research (Dreyer et al., 2010). The relevance of the publications as the historical roots of this newly emerging research field was highlighted in this review. The review cites the four publications in Table 4 and also two further publications with less pronounced peaks (but no other publications published before 1960 which were not identified in our study). The two publications of Schafhaeutl (1840a; 1840b) cited additionally in the review can be seen as precursors to Brodie's publications (Brodie, 1859; 1860). One publication by Schafhaeutl (1840a) appeared in a German journal where fewer citations can be expected.

Insert Table 4 here

The question arises at which point in time the historical publications were cited most frequently. Are such publications already taken account of at the starting point of a new research field (in the case of graphene research this is 2004) since the research is directly based on them? Or are they detected, for example, as forerunners not before literature reviews are published (which discuss the historical background)? Figure 3 shows the evolution over time (citation history) of the four most frequently cited historical publications mentioned above against the backdrop of the time curve for the literature on graphene in total. The citation numbers of the four publications are limited to citing publications dealing with graphene.

Insert Figure 3 here

Only the publication by Wallace (1947) was cited more frequently immediately after the discovery of graphene in the year 2004, whereas the other three historical publications (Brodie, 1859/60; Staudenmaier, 1898; Hummers & Offeman, 1958) did not receive a boost until two years later. This can be explained by the fact that the boom in graphene research was triggered by a physical preparation method and the focus initially was on the physical properties predicted in theory. Accordingly, as a theoretical physics publication, Wallace's publication (1947) was immediately cited more frequently. Over 85% of the citing publications of this paper are classified as physics research. Researchers into chemistry only subsequently started looking at the question of how graphene could be synthesized chemically, which made the other historical publications (Brodie, 1859/60; Staudenmaier, 1898; Hummers & Offeman, 1958) on graphite oxide relevant. Around 70% of the citing publications of these papers are from research into chemistry. A comparison of all the literature on graphene in these two research areas shows that generally speaking, the major reaction of the chemistry community to the discovery of graphene came two years after that of the physics community.

As described above, the discovery of free-standing graphene goes back to the publications by Novoselov et al. (2004). The earliest references in these publications are from the 1980s (1981). One possible reason for the absence of historical publication data could be that the publications are relatively short and focus on the discovery of free-standing graphene. Furthermore, the physical and chemical proofs for the new discovery were given priority. The authors did not discuss the history of the discovery until three years later (Geim & Novoselov, 2007).

A question that can be raised when applying this method is the metaphor of historical roots. Saying that something has a history suggests a chain of causally connected events, i.e. in this case that ideas presented in previous papers are causally connected to ideas presented in later papers. One could see this as a chain of events in which previous papers are necessary and/or sufficient conditions for later publications. Two possible interpretations of Figure 3 are that the historical papers on graphene research shown in Table 4 were re-discovered (if there in fact is some long and invisible chain of events), or that this is historical reconstruction by authors who searched and constructed historical roots, but these reconstructed roots did not really condition the discovery of graphene. It is not possible to give an answer to this question using our method because RPYS traces the referencing behavior of the citing authors without the option of making this distinction.

The papers by Staudenmaier (1898) and Hummers et al. (1958), for example, deal with the preparation of graphite oxide, which has played a major role as a precursor material in efforts aiming at the chemical preparation of graphene. From this perspective, these papers are not the direct historical roots of the material, but they have been most important for current and past graphene research. In contrast, the theoretical analyses by Wallace (1947) suggested that such carbon layers might exhibit extraordinary electronic characteristics, which obviously stimulated the synthesis of the material. The discoverers of the material cited the Wallace paper in their review (Geim & Novoselov, 2007). As a consequence, the RPYS method reveals the historic papers (potentially) most relevant for the evolution of a specific research field which should be taken into consideration when discussing its history. But their specific role can only be determined by careful analysis by experts in the relevant field.

**Example 2: Solar Cells**

Since global warming has turned out to be a serious problem which is already anticipated for the near future, regenerative energy sources (solar energy, wind and water power) have attracted increasingly more attention in science and politics. The field of solar cells / photovoltaic research emerged in the 1980s, stimulated by government programs of the time (Leydesdorff & Van der Schaar, 1987). The scale changed around 2000 when research on solar cells became a major research topic with strongly increasing publication output (both with regard to articles and patents). Its time evolution is quite similar to the time curve of graphene literature presented in Figure 3.

For the subject-specific visualization of research on solar cells we downloaded the set of papers with the string "TI= solar cell OR TI=solar cells OR TI=photovoltaic OR TI=photovoltaics" from WoS on February 7, 2013 (see Table 5). The retrieval result was 22,204 records. This data was used as input to the journal mapping using the routine provided by Leydesdorff, Rafols, and Chen (in press). The explanation of the mapping method is the same as for the graphene example above.

Insert Table 5 here

Insert Figure 4 here

Figure 4 shows that, similar to graphene research, the solar cell research field also stretches throughout large areas of the natural sciences. However, this research field has a strong focus on one specific journal: "Solar Energy Materials and Solar Cells".

For the RPYS analysis, all cited references (1,375,860) have been selected from the complete set of the papers on solar cells published since 1974 (62,412) in the SCI source journals (date of the literature search: Feb. 07 2013). The most frequently cited historic papers have been revealed and displayed as in Figures 2a-2d of the graphene example.

The distribution of the number of references cited in solar cell papers against their publication years is presented in Figures 5a-5d: Figure 5a shows the distribution of the number of all the references cited in solar cell publications across their publication years. Again, the RPYs are presented back to the year 1800. Figure 5b shows a cutout limiting the RPYs to the time period 1800 to 1970 with the distinct peaks more clearly visible. Figures 5c and 5d show the deviation of the number of cited references in one year from the median for the number of cited references in the two previous, the current and the two following years. While Figure 5c shows the absolute deviation from the median, Figure 5d illustrates the deviation in percent. Similar to Figure 2c and 2d, the absolute deviation from the median given in Figure 5c accentuates the peaks of the most frequently cited post-1950 publications whereas the deviation expressed in percent shown in Figure 5d emphasizes the pre-1900 historical publications. Towards the present, the peaks of individual publications lie over a broad continuum of newer publications and the proportion of individual, much-cited publications in the RPYs falls steadily. Therefore, the time scale in Figures 5b-5d is limited to 1970 as the most recent RPY.

Insert Figure 5 here

According to Figure 5c, the most frequently cited RPYs are 1839, 1938, 1952, 1954, and in particular 1961. Figure 5d reveals the RPYs 1806, 1847, and in particular 1839 as the publication years of potentially important historical works. Since Figure 5d accentuates the pre-1900 publications, it also includes here cited papers with field-specific citation counts below 10: the RPY peaks in 1806 (mainly resulting from citations of a paper by C.J.T. De Grotthuss) and 1847 (mainly resulting from a paper by R. Kohlrausch, which has been cited erroneously – see Cardona et al., 2007). Both papers have been omitted here due to their comparatively low citation impact within the solar cell research field (see the citation count limit mentioned in the methods section).

The two other most frequently cited early papers (Becquerel, 1839 a; Becquerel, 1839 b) deal with the world's first photovoltaic system and hence are obviously very important historical papers within the research field analyzed here. The three papers corresponding to the RPYs 1938, 1952, and 1954 address basic phenomena associated with photovoltaics: the frequently investigated Onsager theory as a model for carrier recombination (Onsager, 1938), the Shockley-Read-Hall (SRH) recombination process (Shockley & Read, 1952), and the Burstein-Moss effect (Burstein, 1954). According to Figure 5b and in particular Figure 5c, the 1961 peak is most clearly pronounced. This peak can be assigned to the Shockley-Queisser (SQ) paper (Shockley & Queisser, 1961). The SQ paper belongs to the very few papers that were not noticed throughout many decades or even half a century and were then heavily cited. Such papers, particularly if the time delay and the subsequent boost are clearly pronounced, are called ''sleeping beauties'' (Van Raan, 2004).

The solar cell example shows that there may be other peaks that are less clearly pronounced, but worth analyzing more closely. It is a question of time and effort how many peaks are analyzed later on. These analyses are often complicated, since most of the pre-1900 historical papers are not covered by literature databases. Table 6 specifies the most frequently cited early (pre-1970) papers including their citation data and our short comments concerning their importance for the solar cell / photovoltaic research field.

Insert Table 6 here

All the publications listed in Table 6 are most important for research around solar cells: The Becquerel papers (Becquerel, 1839 a; Becquerel, 1839 b) introduced the photovoltaic process. The two following papers (Onsager, 1938; Shockley & Read, 1952) deal with carrier recombination (the recombination process of electrons and holes as the carriers of electric charge). The next paper (Burstein, 1954) addresses the band gap as the most important parameter of semiconductors in the SQ model. Finally, the SQ-paper reveals the decisive value that is used to gauge the efficiency of solar cells.

**Discussion**

In this study we proposed RPYS, a bibliometric method with which it is possible to determine the historical roots of research fields and quantify their impact on current research. "If you want to know how science is carried out, then in one way or another, you are going to have to look at the history of science" (Lehoux & Foster, 2012, p. 885). However, one can follow a different route by not looking at the history of science, but at the rewrite by practitioners in terms of citation behavior. The RPYS method is based on an analysis of the frequency with which references are cited in the publications in a specific research field by publication year. The origins show up in the form of more or less pronounced peaks mostly caused by individual historical publications which are cited particularly frequently. As the RPYS can only indicate the possible origins, a second step is required in which experts verify which publications genuinely played a significant part in a research field. When those publications which resulted in a peak are identified, each of them should be reviewed for their significance in the particular research field and what contribution they made. RPYS is a very simple method which can be applied in different disciplines. One method which approaches the quantification of historical events in a way similar to RPYS and which can be used to examine historical events on the basis of very different sources of data and mathematical models was proposed by Turchin (2003) and called cliodynamics (Spinney, 2012).

We used research on graphene and on solar cells to illustrate how RPYS functions and what results it can deliver. Many research fields refer in their literature to historical publications which are cited comparatively frequently and can be investigated. However, sometimes, the methods and topics of a research field are so new (e.g. molecular biology or molecular genetics) that the roots do not extend very far into the past. These should be looked at individually. According to Smith (2012) a method like RPYS can be included in the "newly emerging field of 'historical bibliometrics'" (Holmes, 2012). Smith says that this is a "relatively under-researched area" in which new studies would be very welcome. For example, it would be possible to use RPYS to examine Stigler's Law of Eponomy (Stigler, 1980), which says that "no scientific law is named after its discoverer". In a recent study, Grünbaum (2012) for example, looks at the question – without the help of bibliometrics – of whether Napoleon's theorem really is Napoleon's. Another phenomenon in the history of science that would be interesting for RPYS are multiple independent discoveries (Merton, 1973), whereby it would be possible to use bibliometrics to examine the form in which the relevant historical publications on multiple independent discoveries are cited.

Alternative concepts to the RPYS for analyzing historical papers are the concept of co-citations and research fronts (Garfield, 1993) and also a method called "algorithmic historiography" (Garfield, 2001; Garfield et al., 2003; Leydesdorff, 2010). The HistCite™ software developed by Alexander Pudovkin and Eugene Garfield (http://garfield.library.upenn.edu/algorithmichistoriographyhistcite.html) enables the establishment of the citation graph (sometimes called historiogram or historiograph) which visualizes the citation network among publication sets, including historical papers. The RPYS method proposed here is far more simple than the alternatives and does not reveal the citation network of the historical papers. RPYS only reveals quantitatively which historical papers are of particular interest for the specific research field / topic.

Bibliometrics of historical papers is basically limited by two kinds of under-citation (Garfield, 1975; Marx & Cardona, 2009; McCain, 2011, 2012): "obliteration by incorporation" and the "palimpsestic syndrome." These two phenomena were first described by the sociologist Robert K. Merton (1965). The process of obliteration affects seminal works offering novel ideas that are rapidly absorbed into the body of scientific knowledge. Such work is thereafter integrated into textbooks and becomes increasingly familiar within the scientific community. The palimpsestic syndrome (referring to a piece of parchment that is erased more than once to make room for newer work) means covering over an idea by ascribing it to a more recent author, who cites the original work. As a result of the absorption, canonization, and covering over, the original sources fail to be cited, either as full references (formal citations) or even as names or subject-specific terms (informal citations, implicit citations).

**Conclusions**

The RPYS method presented here enables the detection of the most-frequently cited historical papers cited within the field-specific literature and the quantification of their citation impact on more current research. These papers normally comprise the historical roots of the corresponding research field. However, only careful analysis of the detected papers can reveal their real significance. The seminal historical papers selected out of the candidate list (i.e. the highly-cited historical papers of a specific research field) can be analyzed further by bibliometric methods (e.g. time curves of their citations, co-citations etc.). Alternative methods to RPYS like "algorithmic historiography" focus on the network of publication sets (including the early papers), whereas RPYS mirrors the reference side. The decisive advantage of the RPYS method is that the historical papers are detected on the basis of the references cited by the relevant community and without any further assumptions. The direct selection of such papers via an appropriate search query is hardly possible. The papers comprising the historical roots of current research normally deal with more unforeseeable aspects. These aspects can be identified by RPYS. For example, the historical graphene papers deal with graphite or graphite oxide and not with graphene itself.

Acknowledgement: We are grateful to Malcolm Walker, Royal Meteorological Society (UK), Chairman of the History of Meteorology and Physical Oceanography Special Interest Group. He raised the question "Which old papers have been cited most in meteorology?" (Walker, 2010) and thereby stimulated the development of the RPYS method.

**Tables and Figures**

**Table 1:** Data for the visualization of research on Graphene.

|  | "Graphene" |
|---|---|
| Downloads from WOS (SCI-E, SSCI, A&HCI) | 16,145 |
| Journal titles matched with JCR 2011 | 668 |
| Records included in the mappings | 15,895 |
|  | (98.5%) |
| Rao-Stirling diversity | 0.072 |
| WOS Categories attributed to the sets | 32,200 |

**Table 2:** Search query for the RPYS of the literature on graphene.

```
=> dis hist

      (FILE 'HOME' ENTERED AT 15:13:03 ON 07 FEB 2013)

      FILE 'SCISEARCH' ENTERED AT 15:13:13 ON 07 FEB 2013
CHARGED TO COST=IVS
L1         23443 S GRAPHENE#
               SET TERM L#
L2           SEL L1 1- RPY :     190 TERMS

=> dis l2 1- alpha
L2           SEL L1 1- RPY :     190 TERMS

TERM #   # OCC  # DOC  % DOC RPY
------ ------- ------ ------ ---------------
   ...
   36      4      2   0.01 1850
   37      2      2   0.01 1852
   38      5      5   0.02 1853
   39      3      2   0.01 1854
   40     17     17   0.07 1855
   41      1      1   0.00 1856
   42      4      4   0.02 1857
   43      1      1   0.00 1858
   44    156    156   0.67 1859
   45    102    102   0.44 1860
   46      2      2   0.01 1861
   47      3      3   0.01 1865
   48      7      5   0.02 1866
   49      5      5   0.02 1867
   50      1      1   0.00 1870
   ...
```

**Table 3:** Search query for the cited references in 1859/60.

```
=> dis hist

     (FILE 'HOME' ENTERED AT 15:13:03 ON 07 FEB 2013)

     FILE 'SCISEARCH' ENTERED AT 15:13:13 ON 07 FEB 2013
CHARGED TO COST=IVS
L1        23443 S GRAPHENE#
             SET TERM L#
L2          SEL L1 1- RPY :     190 TERMS
L3          254 S L1 AND (1859 OR 1860)/RPY
L4          SEL L3 1- RE HIT :      15 TERMS

=> dis l4 1- occ
L4          SEL L3 1- RE HIT :      15 TERMS

TERM #   # OCC  # DOC  % DOC  RE
------  -------  ------  ------  ---------------
     1     145    145   57.09  BRODIE B C, 1859, V149, P249, PHILOS T ROY SOC LONDON
     2      89     89   35.04  BRODIE B C, 1860, V59, P466, ANN CHIM PHYS
     3       6      6    2.36  BRODIE M B C, 1860, V59, P466, ANN CHIM PHYS
     4       3      3    1.18  BRODIE B C, 1859, V10, P249, P ROY SOC LONDON
     5       3      3    1.18  BRODIE B, 1859, V149, P249, PHILOS T R SOC LONDON
     6       2      2    0.79  BRODIE B C, 1859, V10, P249, P R SOC LONDON
     7       2      2    0.79  BRODIE B C, 1860, V12, P261, Q J CHEM SOC
     8       1      1    0.39  BRODIE B C, 1859, V10, P11, P R SOC LONDON
     9       1      1    0.39  BRODIE B C, 1859, V149, P10, PHILOS T R SOC
    10       1      1    0.39  BRODIE B C, 1860, V114, P6, LIEBIGS ANN CHEM
    11       1      1    0.39  BRODIE B, 1860, P59, ANN CHIM PHYS
    12       1      1    0.39  BRODIE B, 1860, V59, P17, NN CHIM PHYS
    13       1      1    0.39  BRODIE B, 1860, V59, P7, ANN CHIM PHYS
    14       1      1    0.39  BRODIE E C, 1860, V59, P466, ANN CHIM PHYS
    15       1      1    0.39  BRODIE F R S, 1859, V149, P249, PHILOS T R SOC LONDON
********  END OF L4 ***
```

**Table 4:** The four most frequently cited early (pre-1990) references in graphene literature. In each case, the relevant RPY, the number of references in the graphene literature attributed to the specific publication, the total number of references in the graphene literature with regard to the given RPY, the overall number of citations of the specific publication until February 2013 (TC=Times Cited), and the relevant comment from Dreyer et al. (2010) are listed (date of the citation search: Feb. 07 2013).

| RPY | Reference / Comment | TC |
|---|---|---|
| | 204 of 209 references refer to: | |
| 1859 | Brodie, B.C. (1859). On the atomic weight of graphite. Philosophical Transaction of the Royal Society of London, 149, 249-259. | 324 |
| 1860 | Brodie, B.C. (1860). Sur le poids atomique du graphite [On the atomic weight of graphite]. Annales de Chimie et de Physique, 59, 466-472. | |
| | *"In 1859, the British chemist Brodie used what may be recognized as modifications of the methods described by Schafhaeutl in an effort to characterize the molecular weight of graphite by using strong acids (sulfuric and nitric), as well as oxidants, such as KClO3" (p. 9337).* | |
| | 177 of 183 references refer to: | |
| 1898 | Staudenmaier, L. (1898). Verfahren zur Darstellung der Graphitsäure [Method for the preparation of graphitic acid ]. Berichte der Deutschen Chemischen Gesellschaft, 31, 1481-1487. | 270 |
| | *"Nearly 40 years later, Staudenmaier reported a slightly different version of the oxidation method used by Brodie for the preparation of GO by adding the chlorate salt in multiple aliquots over the course of the reaction instead of in a single portion" (p. 9338).* | |
| | 962 of 1085 references refer to: | |
| 1947 | Wallace, P.R. (1947). The band theory of graphite. Physical Review 71, 622-634. DOI: 10.1103/PhysRev.71.622 | 1467 |
| | *"As early as the 1940s, a series of theoretical analyses suggested that these layers—if isolated—might exhibit extraordinary electronic characteristics (e.g., 100 times greater conductivity within a plane than between planes). About 60 years later, these predictions were not only proven correct, but the isolated* | |

*layers of graphite were also found to display other favorable properties …" (p. 9336).*

2095 out of 2971 references refer to:

1958    Hummers , W.S. (Jr.) & Offeman, R.E. (1958). Preparation of graphite oxide.    2511
        Journal of the American Chemical Society, 80(6), 1339-1339. DOI:
        10.1021/ja01539a017

*"Graphite oxide: A berthollide layered material prepared by treating graphite with strong oxidants, whereby the graphite surface and edges undergo covalent chemical oxidation. The degree of oxidation may vary, though strongly oxidized graphite oxide typically exhibits a C/O ratio of approximately 2:1" (p. 9342).*

**Table 5:** Data for the visualization of research on solar cells.

|  | "solar cell(s)" OR "photovoltaic" |
|---|---|
| Downloads from WOS (SCI-E, SoSCI, A&HCI) | 32,251 |
| Journal titles matched with JCR 2011 | 724 |
| Records included in the mappings | 28,398 (88.1%) |
| Rao-Stirling diversity | 0.063 |
| WOS Categories attributed to the sets | 44,651 |

**Table 6:** The six most frequently cited early (pre-1990) references within solar cell literature. In each case, the relevant RPY, the number of references within the solar cell literature assigned to the specific paper, the total number of references within the solar cell literature with regard to the given RPY, the overall number of citations (TC=Times Cited) of the specific paper, and a short comment for explanation of the content are listed (date of the citation search: Feb. 07 2013).

| RPY | Reference / Comment | TC |
|---|---|---|
| | 63 out of 66 references refer to: | |
| 1839 | Becquerel, A.E. (1839). Mémoire sur les effets électriques produits sous l'influence des rayons solaires [ ]. Compt. Rend. Acad. Sci. 9, 561-567. | 165 |
| 1839 | Becquerel, A.E. (1839). Recherches sur les effets de la radiation chimique de la lumiere solaire au moyen des courants electriques [ ]. Cr. Hebd. Acad, Sci. 9, 145-149. | 73 |
| | *Alexandre-Edmond Becquerel (1820-1891) built the world's first photovoltaic system. In this experiment, silver chloride in an acidic solution was illuminated while connected to platinum electrodes, thus generating photovoltage and photocurrent.* | |
| | 130 out of 250 references refer to: | |
| 1938 | Onsager, L. (1938). Initial recombination of ions. Physical Review 54(8), 554-557. DOI: 10.1103/PhysRev.54.554 | 1356 |
| | *"A number of different models have been used to explain carrier recombination in low mobility materials. The most frequently investigated model has been the Onsager theory of geminate recombination [72]. In this theory it is assumed that electron-hole pairs separate to an initial distance $r_0$ in the primary photogeneration step" (Chamberlain, 1983; p. 63-64).* | |
| | 353 out of 813 references refer to: | |
| 1952 | Shockley, W., Read, W.T. (1952). Statistics of the recombinations of holes and electrons. Physical Review 87(5), 835-842. DOI: 10.1103/PhysRev.87.835 | 2911 |
| | *This is the original paper on the so-called Shockley–Read–Hall (SRH) process/recombination: a two-step recombination process where conduction electrons can relax to the defect level and then relax to the valence band,* | |

*annihilating a hole.*

205 out of 1133 references refer to:

1954    Burstein, E. (1954). Anomalous optical absorption limit in InSb. Physical    1656
Review 93(3), 632-633. DOI: 10.1103/PhysRev.93.632

*This is the original paper of the so-called Burstein–Moss effect/shift: the
increase of the bandgap of semiconductors as the absorption edge is pushed
to higher energies due to doping effects.*

923 out of 2349 references refer to:

1961    Shockley, W., Queisser, H.J. (1961). Detailed balance limit of efficiency of p-n    1173
junction solar cells. Journal of Applied Physics 32(3), 510-519. DOI:
10.1063/1.1736034

*This paper deals with the energy conversion efficiency of solar cells which is
the percentage of power converted from sunlight to electrical energy. The so-
called Shockley–Queisser efficiency limit (also named Shockley–Queisser limit
or detailed balance limit) refers to the maximum theoretical efficiency of solar
cells using a single p-n junction to collect power (the p-n junction refers to the
boundary of two semiconductors, the p-type and the n-type; the p-type
semiconductor contains excess holes while the n-type contains excess free
electrons as the carriers of electric charge). The Shockley-Queisser limit puts
the maximum solar cell efficiency at about 34%. This means that at most, only
34% of sunlight can be converted into electrical energy. Currently, silicon-
based photovoltaic cells have an efficiency of around 22%. However, cells with
multiple layers (tandem cells) can outperform this limit.*

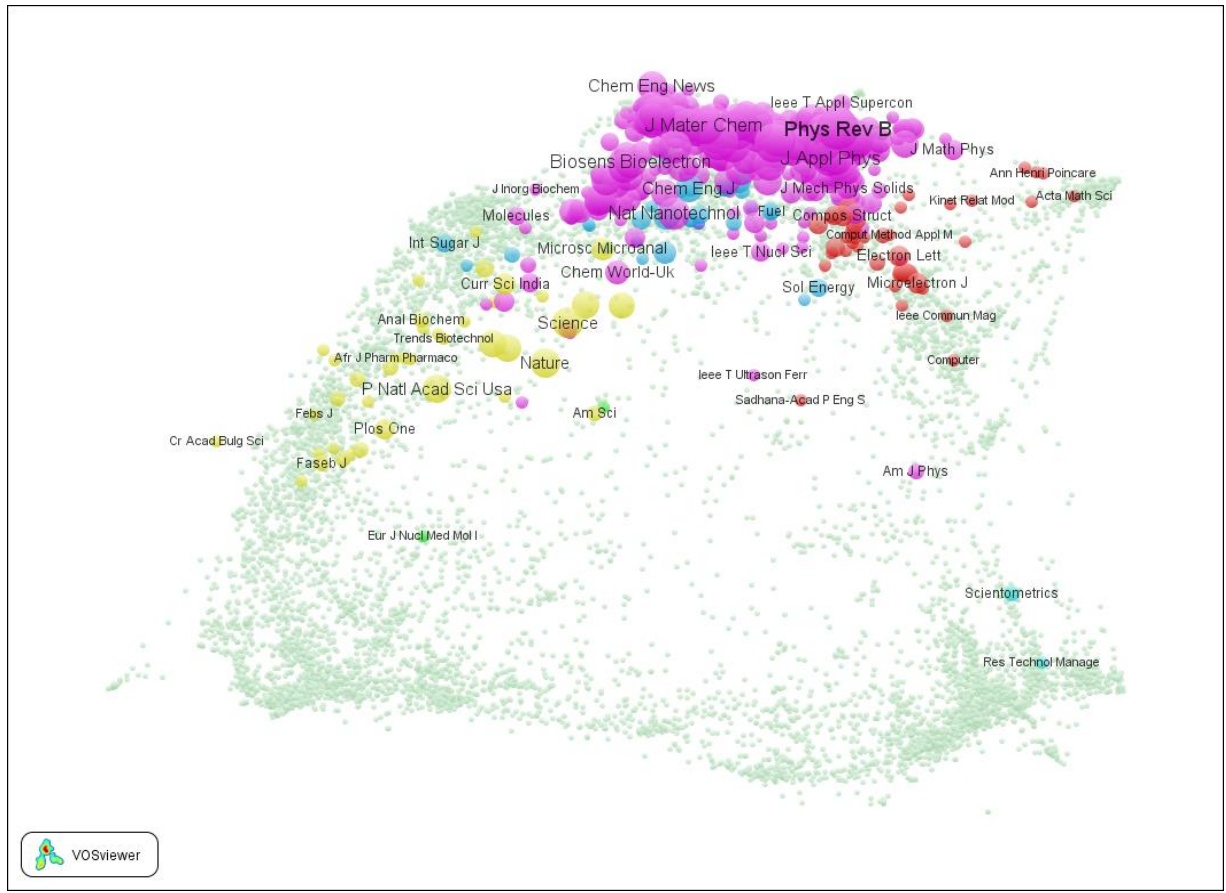

**Figure 1:** Journal mapping of 668 journals containing 15,895 papers with the word "graphene" in the title. This map can also be web-started at

http://www.vosviewer.com/vosviewer.php?map=http://www.leydesdorff.net/graphene/graphene.txt.

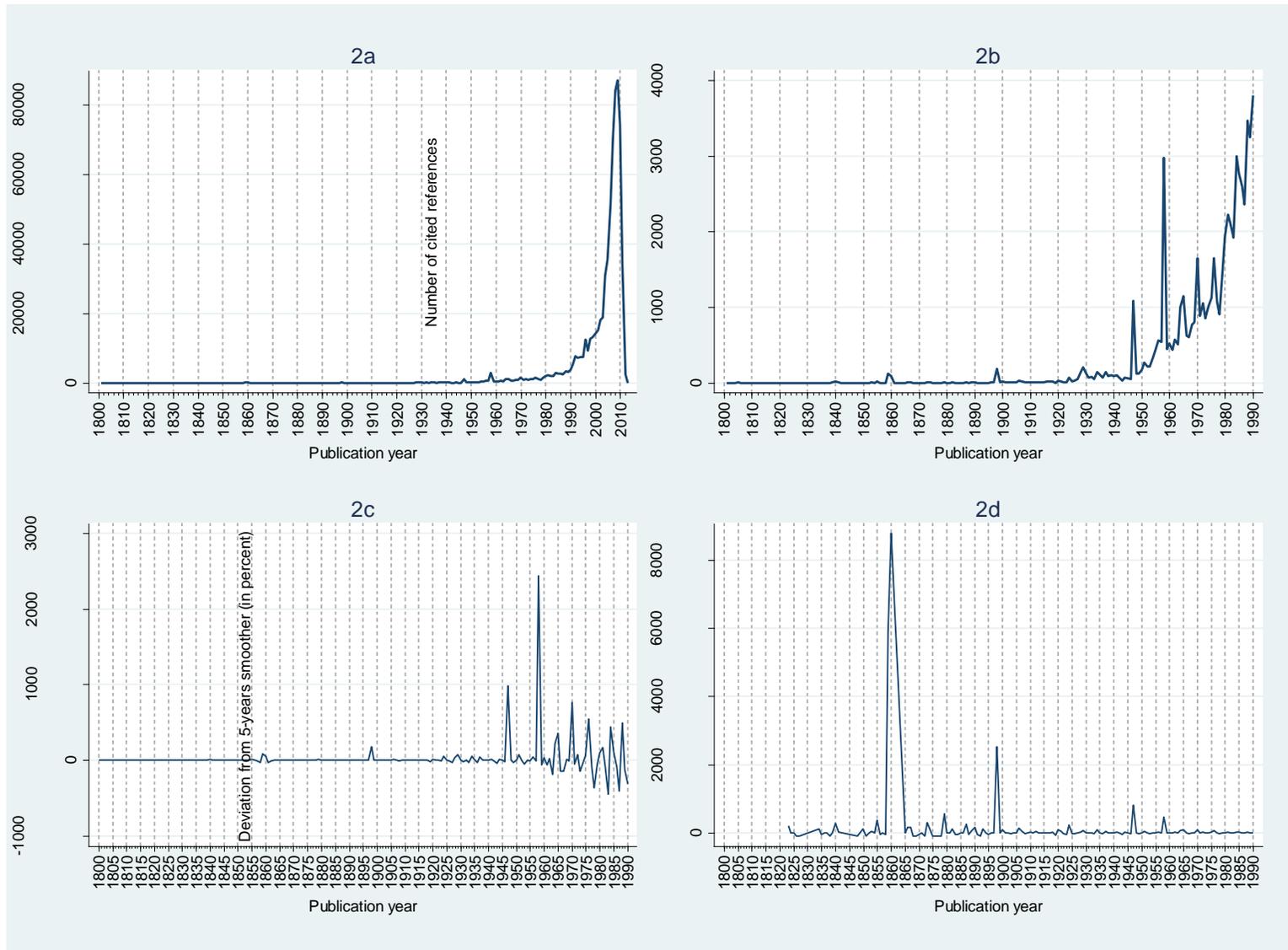

**Figure 2 a-d:** Annual distributions of cited references in publications of research on grapheme.

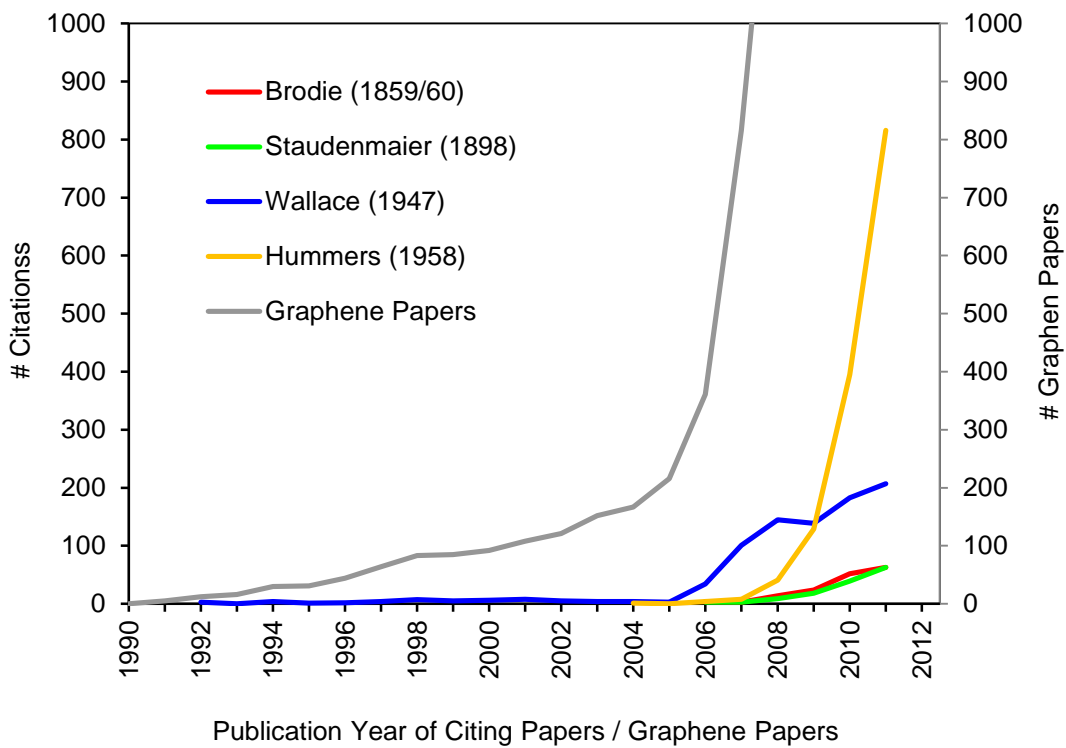

**Figure 3:** Citation history of the four most frequently cited historical publications in graphene literature.

The overall number of graphene papers published per year is shown for comparison.

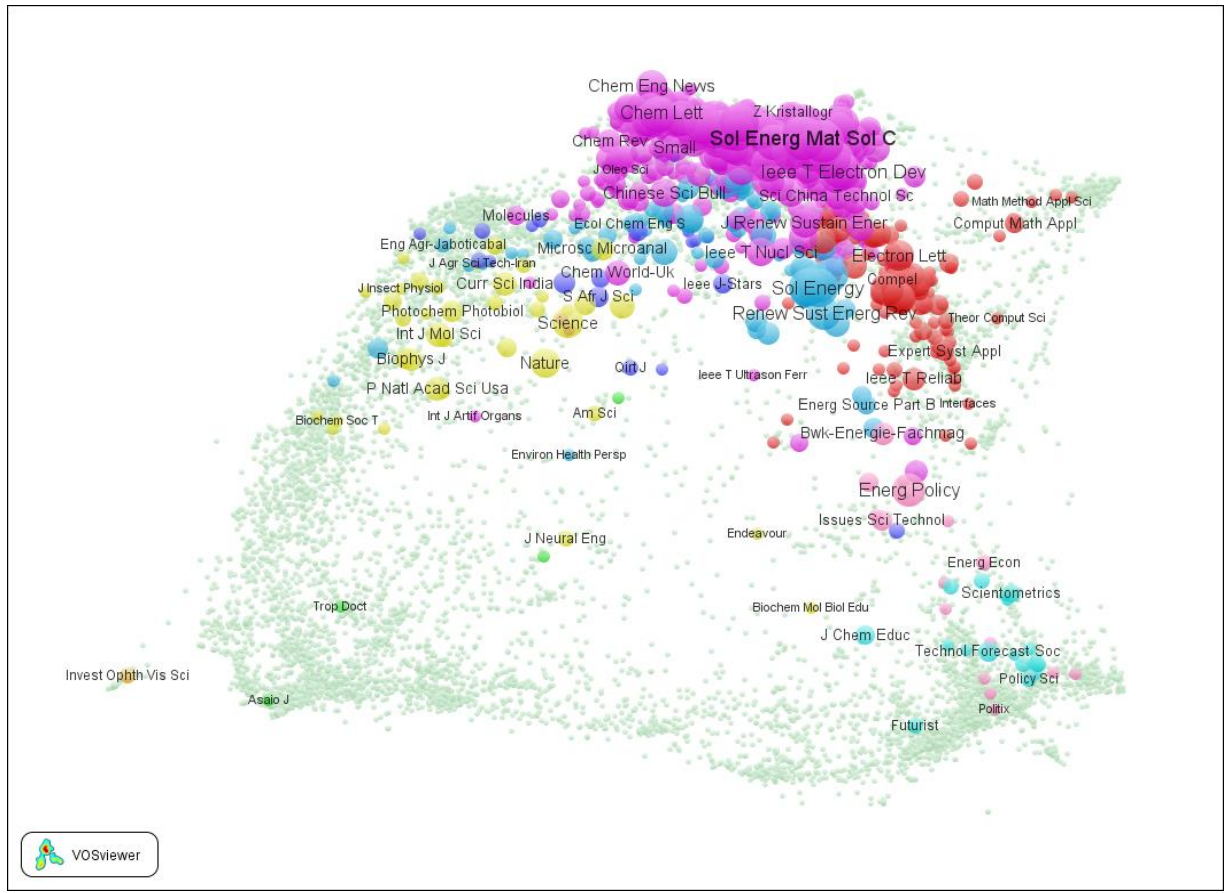

**Figure 4:** Journal mapping of 724 journals containing 19,824 papers with "solar cell(s)" or

"photovoltaic(s)" as words in the title. This map can also be web-started at

http://www.vosviewer.com/vosviewer.php?map=http://www.leydesdorff.net/graphene/solarcells.txt&view=2.

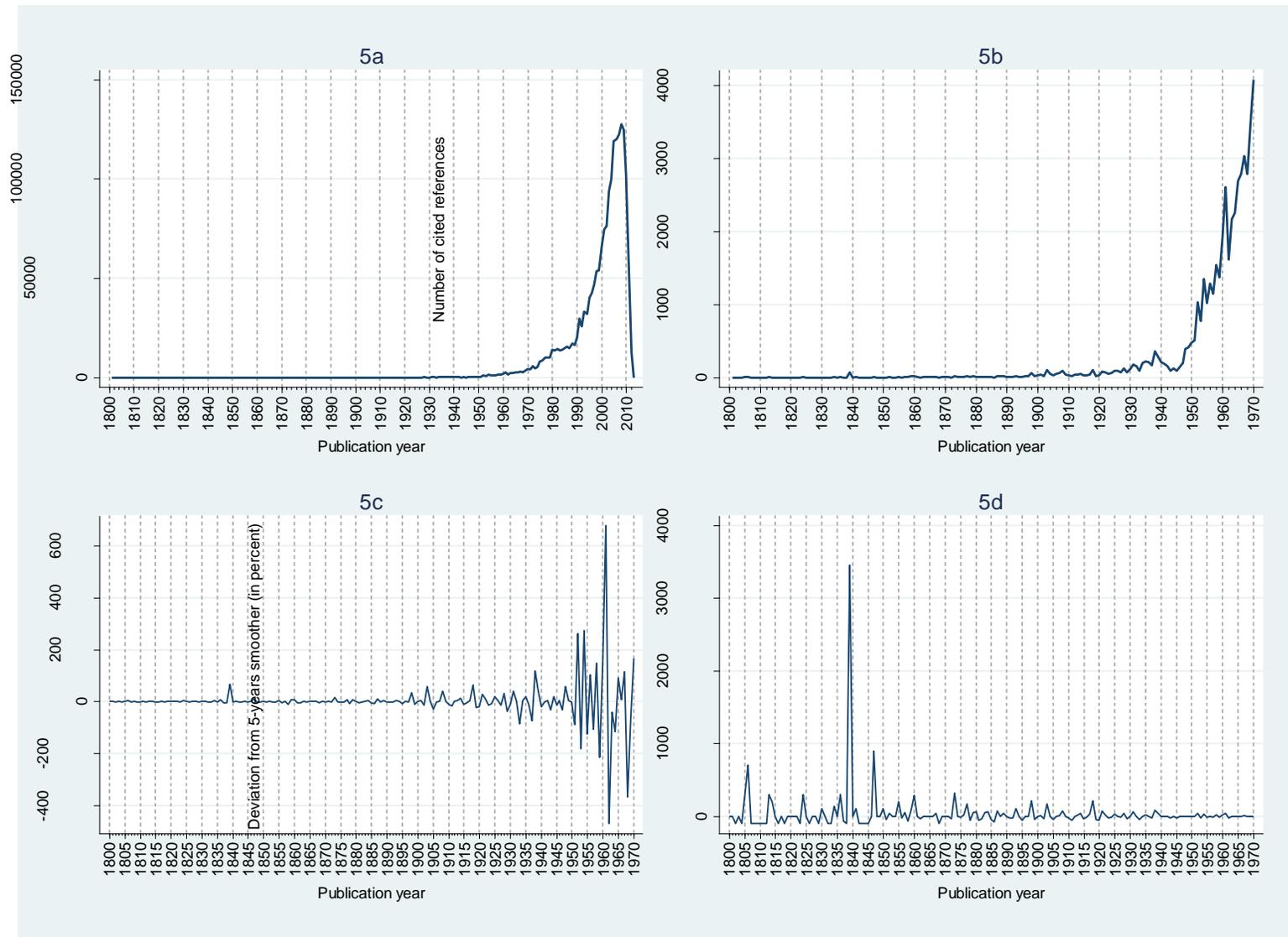

**Figure 5 a-d:** Annual distributions of cited references in publications of research on solar cells.